# The progression of visual search in multiple item displays: First relational, then feature-based.

Zachary Hamblin-Frohman, Koralalage Don Raveen Amarasekera & Stefanie I. Becker

School of Psychology, The University of Queensland, Brisbane, Australia.

**Abstract**

It is well-known that visual attention can be tuned in a context-dependent manner to elementary features, such as searching for all *redder* items or the *reddest* item, supporting a *relational theory* of visual attention. However, in previous studies, the conditions were often conducive for relational search, allowing successfully selecting the target relationally on 50% of trials or more. Moreover, the search displays were often only sparsely populated and presented repeatedly, rendering it possible that relational search was based on context learning and not spontaneous. The present study tested the shape of the attentional tuning function in 36-item search displays, when the target never had a maximal feature value (e.g., was never the reddest or yellowest item), and when only the target colour but not the context colour was known. The first fixations on a trial showed that these displays still reliably evoked relational search, even when participants had no advance information about the context and no on-task training. Context learning further strengthened relational tuning on subsequent trials, but was not necessary for relational search. Analysing the progression of visual search within a singe trial showed that attention is first guided to the relationally maximal item (e.g., reddest), then the next-maximal (e.g., next-reddest) item, and so forth, before attention can hone in on target-matching features. In sum, the results support two tenets of the relational account, that information about the dominant feature in a display can be rapidly extracted and used to guide attention to the relatively best-matching features.

**Introduction**

It is well-known that we cannot consciously process all objects in a visual scene at once. To address this, visual attention selects objects for in-depth processing, often guiding our gaze to relevant parts in a scene (e.g., Deubel & Schneider, 1996). Much effort has been devoted to determine which items in a scene will be attended first, and more generally, to identify the processes involved in creating our rich mental representation of the visual environment (for a review, see Carrasco, 2011; Wolfe, 2021).

To date, it is widely accepted that attention can be guided by both, bottom-up, stimulus-driven processes and top-down, goal-driven processes (e.g., Wolfe, 2020). For example, attention can be reflexively drawn to visually salient events such as a bright flash, a movement, or the sudden appearance of an object (e.g., Theeuwes, 2004, 2013), or it can be top-down tuned to select items with certain attributes (e.g., colours: red, green) to help goal-related behaviours such as finding a friend in a crowd (e.g., Desimone & Duncan, 1995; Wolfe, 1994, 2021) . Correspondingly, current models of visual attention typically include both a bottom-up and a top-down component to predict which items in a visual scene will be selected first (e.g., Wolfe, 1994, 2021). Top-down tuning is typically modelled as an increase or decrease in the firing rate of sensory neurons in response to specific stimulus attributes (e.g., red, green; Navalpakkam & Itti, 2007; Yu, Hanks & Geng, 2022). For example, when looking for an orange in a fruit basket, we would tune attention to orange, which increases the output of neurons that respond to orange and prioritises colour-matching items for selection.

It is commonly assumed that attention is tuned to the feature value that a person is looking for (e.g., particular shade of orange; e.g., Duncan & Humphreys, 1989; Navalpakkam & Itti, 2007). However, to date, there are also several accounts of non-veridical tuning. Navalpakkam and Itti (2007) noted that tuning attention to the exact target feature value would not be beneficial when the target is very similar to surrounding irrelevant non-target items, as tuning attention to, for example, orange, would also boost the response gain of red-orange or yellow-orange, leading to a poor signal-to-noise ratio (SNR). They proposed that attention would be tuned to a feature value that is slightly shifted away from similar nontargets, to increase the SNR (e.g., to yellow-orange, when an orange target is presented among red-orange items; Navalpakkam & Itti, 2007). According to their *optimal tuning account*, attention is always tuned to the feature value that maximises the ability to discriminate the target from the non-targets (i.e., maximise the SNR). Thus, attention would only be tuned to the exact target feature value when the target is presented alone or among dissimilar other items. In less discriminable cases, attention should be shifted to a slightly exaggerated target feature value. In line with this idea, a perceptual probe task revealed that a slightly shifted non-target colour was likely to be mistaken for the target when the target was consistently embedded among similar featured non-targets in a prior visual search. For example, if an orange target was always presented among similar, yellow-orange nontargets in a visual search task, participants would pick a slightly redder (red-orange) colour as the target colour in intermixed probe trials (Navalpakkam & Itti, 2007; see also Geng et al., 2017; Scolari et al., 2012).

Another account of non-veridical tuning is the Relational Account, which proposes that attention may not at all be tuned to a specific feature value. As noted by Becker (2010), tuning attention to a particular feature value may not be beneficial in natural environments, as specific feature values such as the size, shape and colour of objects vary a lot with differences in distance, perspective and shading. To allow efficient selection of the targets in these noisy environments, attention could be tuned in a context-dependent manner to objects, biasing attention to a feature that the target has *relative* to the other items in the context (e.g., redder, larger, darker; Becker, 2010). For example, when searching for an orange in a fruit basket, the visual system would quickly assess the distribution of colours in the visual scene to determine the dominant colours, and tune attention to the relative colour that would best discriminate the target from the dominant coloured non-target items. Thus, attention would be tuned to all redder items or the reddest item when the fruit basket (or visual scene) contains many yellow or green objects, and tune attention to all yellower items or the yellowest item when the fruit basket (or visual scene) contains many red objects. As a consequence of this very broad tuning, the item that maximally fulfils the relevant feature relationship will be selected first (e.g., reddest or yellowest item). Hence, when attention is tuned to all redder items, the reddest item in the visual field will be selected first, followed by the next-reddest, and so forth.

In line with this prediction, several visual search studies showed that when an orange target is presented among mostly yellow(er) items, a red irrelevant distractor was more likely to be attended first than the target, even when it was quite dissimilar from the target, suggesting that attention was biased to all redder items, or the reddest item (e.g., Becker, 2010; Becker et al., 2013; Hamblin-Frohman & Becker, 2021; York & Becker, 2020). Selection of the distractor was reflected both in higher response times (RT) to the target, as well as in a high proportion of first eye movements to the red distractor. Several studies also included a visually salient distractor with a dissimilar colour (e.g., blue), but found no or only very weak effects of saliency, ruling out that selection of the relatively matching (e.g., red) distractor was mediated by bottom-up, stimulus-driven processes (e.g., Martin & Becker, 2018; York & Becker, 2020).

Both optimal tuning and the relational account propose that attention can be tuned to features in a non-veridical manner but propose very different underlying mechanisms. Subsequent studies combined the two paradigms to determine whether attention is tuned to all relatively matching features or an optimally shifted feature value (see Figure 1 for an illustration of the different accounts).

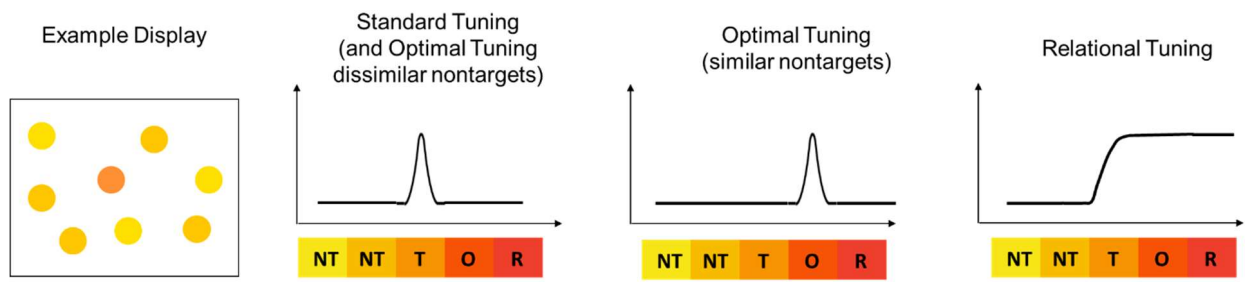

*Figure 1*. Overview of different theoretical tuning functions to tune attention to the target in the example display (left), which shows an orange target (T) among yellow-orange and yellow non-targets (NT). The standard view (2nd from left) is that attention would be tuned to the target colour (T). According to Optimal Tuning (3rd from left), attention would be tuned to a slightly exaggerated, 'optimal' target feature value that is shifted away from the nontargets, to increase the SNR (optimal colour; O). According to the Relational Account (right), attention would be tuned to the relative colour of the target, that the target has relative to the other items in the surround (here: redder), which can include vastly dissimilar colours (relatively matching colour, R).

The results of two studies showed that attention is biased to all relatively matching items, indicating that early visual selection (i.e., which item is selected first) follows the predictions of the relational account. However, target identification judgements were skewed towards the slightly shifted feature value, showing that perceptual decision-making (i.e., decisions about whether a selected item is the target or not) is best described by the optimal tuning account (Hamblin-Frohman & Becker, 2021; Yu et al., 2022). Thus, visual attention can be tuned rather broadly towards the target's relative feature, leading to frequent selection of very dissimilar colours that only share the target's relative colour, whereas perceptual decision-making is tuned rather sharply towards an exaggerated target feature value, so that only very similar colours are mistaken for the target colour.

Despite the progress in describing the attentional tuning function, there are still significant knowledge gaps. First, one important claim of the relational account is that the visual system quickly assesses the dominant feature in a scene and biases attention to the relative feature that discriminates the target from the (dominant feature in the) context. However, this fast, automatic evaluation of the context has never been tested, as the target and non-targets were typically repeated numerous times in previous studies, and the reported selection behaviour was based on average trial performance (e.g., Becker, 2010). This leaves open the possibility that the target's discriminative relative feature must be learnt, or that the visual system adapts weight settings over consecutive trials in a trial-and-error fashion to bias attention to the relative feature – which is very different from the claim that attention can be biased to relative features instantly, with the first glance.

Second and relatedly, the boundary conditions for tuning to relative features are currently not clear. Previous work has shown that attention will be tuned to the relative target colour when the target is the reddest (or greenest / yellowest / bluest) item in the display on ≥50% of the trials, including when the target always has the same feature and could be found by tuning attention to the exact feature value (e.g., Becker et al., 2013). Relational search was also observed across different feature dimensions (e.g., colour, size, brightness, shape), and across different search tasks such as pop-out search (e.g., Becker et al., 2014), feature search (e.g., Becker et al., 2013) and conjunction search (e.g., Becker et al., 2017). However, it is currently unknown if we would observe relational search when the target is never the reddest (or greenest / yellowest / bluest) item in the display, but differs from the majority of non-target items in one direction (e.g., redder).

Previous studies have shown that we can enforce tuning to a specific feature value; for example, by presenting a target among equal numbers of flanking nontarget colours, such as presenting an orange target among three yellow and three red nontargets (Becker et al., 2014). As the target is neither the reddest nor the yellowest item, attention cannot be tuned to the relative feature but needs to be tuned to the target's feature value in these displays (e.g., Becker, Harris, Venini & Retell, 2014; Harris, Remington & Becker, 2013; Schoenhammer et al., 2016; but see Becker et al., 2013).

Given the ability to tune attention to specific feature values, it is unknown whether participants would adopt a relational search or feature-specific search when the target is never the reddest item in the visual field, yet differs from a majority of items in a single direction (e.g., redder). With that, it is currently unclear whether relational search is indeed based on a mechanism assessing the dominant feature in the visual field, or whether it is only adopted when relational search allows localising the target successfully on a large proportion of trials (i.e., through statistical learning).

Third, previous work on characterising the attentional tuning function is also limited in that the studies typically used fairly sparse displays, of four to eight items (e.g., Becker, 2010; Hamblin-Frohman & Becker, 2021, Navalpakkam & Itti, 2007; Yu et al., 2022). The results typically showed that visually salient items (e.g., a red item among blue-green stimuli) did not attract attention, or attracted attention only very weakly (e.g., Gaspelin et al., 2015; Martin & Becker, 2018; York & Becker, 2020). These results were taken to show that strong capture by relational items was not mediated by saliency, and that bottom-up saliency may modulate attention to a lesser extent than assumed in current models (e.g., Theeuwes, 2013; Wolfe, 1994). However, Wang and Theeuwes (2020) argued that item salience is drastically weakened in sparse displays. If this is correct, previous studies on attentional tuning may have overestimated the role of tuning to relative features and/or underestimated the role of bottom-up saliency effects in guiding attention. Thus, it is still an open question whether previous results would generalise to more realistic search settings, such as search displays containing a larger number of search items.

**Aims**

The aim of the present study was to test some of the untested assumptions of the relational account in more realistic search settings. To that end, the search displays always contained thirty-six items, one of which was a saliently different distractor. This allowed assessing possible contributions of bottom-up saliency effects to visual selection in more realistic search conditions with potentially stronger saliency effects.

To assess how attention is guided with versus without training on the task, we created four blocked conditions in which the search colours varied (bluer, greener, yellower and redder target; see Fig. 2). Participants completed ten trials in each block, and then switched to a different search colour. Prior to each block, participants were only informed about the exact target colour (teal or orange), but not about the colour of the non-target items or the target's relative colour (redder, yellower, greener, or bluer). Thus, analysing the first trial of each block allows assessing how attention was tuned to the target under conditions of uncertainty and in the absence of training effects.

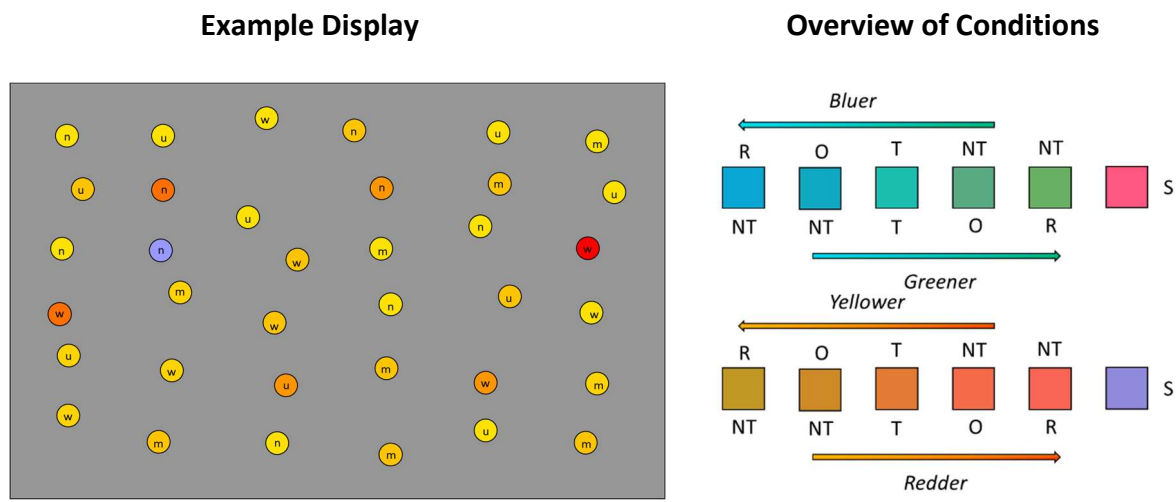

*Figure 2.* Left: Example of a visual search display. Search displays consisted of three target-coloured items (here: orange) presented among 29 nontarget coloured items that could have two different colours (here: yellow-orange or yellow). Participants were instructed to search for a target-coloured item (here: orange) that contained the letter *u* or *m*, and to report whether the target contained the u or m (whereas the other orange items contained the letters *w* or *m*). To assess how attention was guided in search, we assessed eye movements to items with a relationally matching colour (here: red), optimal colour (here: red-orange), and a saliently different colour (here: violet). Right: Colours used to create the four blocked conditions. The target could be teal or orange, and the teal target could be presented among other green-ish non-targets (bluer target), blue-ish non-targets (greener target), while the orange target could be among red-ish non-targets (yellower target) or yellow-ish non-targets (redder target). Prior to a mini-block of 10 trials, participants were informed of the target colour, but not of the colour of the non-targets, which determined whether the target would be relationally redder/yellower or greener/bluer than the non-targets.

Deviating from previous studies, the search displays always contained all possible distractor colours, allowing a more fine-grained measurement of distractor effects and an assessment of possible boundary conditions for relational tuning. For example, in the redder target condition, an orange target was presented among 29 yellow-orange and yellow nontargets, one distractor with a relatively matching colour (red), two distractors with an optimal colour; i.e., an exaggerated target colour that was slightly shifted away from the nontarget colours (red-orange), two distractors with the target colour (orange), and a saliently different distractor with an unrelated colour (e.g., a blue; see Figure 2). These displays allows assessing how attention is tuned to the target when the target differs from the majority of non-targets in a single direction (redder) but is never the relationally maximal item (e.g., reddest, bluest, greenest or yellowest item) in the display.

To assess how attention was tuned to the target, we measured eye movements to each of the different item types (target and distractors). Specifically, to tap into early processes of visual selection, we analysed the first eye movements on a trial, which are not influenced by prior fixations or search processes.

If attention is tuned to the exact target colour, we would expect a large proportion of first eye movements on the target-coloured items (orange, teal), regardless of the context colour, and only few eye movements on the differently coloured distractors (optimal, relational). On the other hand, if attention is tuned to the optimal colour, we would expect most first eye movements on the optimal colour and a decrease in selecting distractors with other-than-optimal colours. According to the relational account, we would expect most first eye movements directed towards the distractor that best matches the relative colour of the target, viz., the distractor that is the reddest, yellowest, greenest, bluest in the visual field, followed by the optimal distractor (as this is the next-maximal, e.g., next-reddest item), and the target-coloured distractors.

Moreover, if relational search is based on a fast, automatic assessment of the dominant colour in the display, we would expect to see these results immediately in the first trial of each block, prior to learning and knowledge of the relative values.

If the salient distractor automatically captures attention in these multiple-item displays (e.g., Wang & Theeuwes, 2020), we would expect a large proportion of first eye movements on the salient distractor. Moreover, if participants can learn to ignore or inhibit the distractor, we would expect this saliency-effect to decrease with training (i.e., over the course of a block; e.g., Gaspelin & Luck, 2018; Hamblin-Frohman et al., 2022).

## Method

**Participants.** To estimate the required sample size, we examined the ability to detect relational vs. feature-specific search in the first fixations on distractors in previous work (Becker, Harris, Venini & Retell, 2014). The weakest effect was the feature-specific effect ($t(14) = 2.5$, p = .024 in search for a redder target; $t(14) = 2.5$, p = .027 in search for a yellower target; Becker et al., 2014; Exp. 3). The BUCSS tool suggested a target sample size of N=32 for the present study, to achieve a power of 85% (with 50% assurance; Anderson et al., 2017).

Thirty-four paid participants from the University of Queensland participated in the experiment. Two participants were excluded for having a low search accuracy (< 70%), leaving 32 in the final analysis (*M* age =23.1 years (*SD* = 1.9), 24 female). The study was approved by the University of Queensland ethics board, and all procedures were in line with the Declaration of Helsinki.

**Apparatus.** Stimuli were presented on a 21-inch CRT monitor with a refresh rate of 85Hz. A chin and headrest were used to hold the participant's heads 600mm from the screen. Gaze location was measured by an SR-Research Eyelink-1000 eye tracker at 500Hz sampling rate. The experiment was controlled by Python's PsychoPy (Peirce, 2007).

**Stimuli.** All stimuli were presented against a grey background. Each search array contained thirty-six coloured circles (radius: 0.48°) arranged in a six-by-six grid format (see Figure 2). Stimulus locations were initially selected to have a 4.96° horizontal separation and 4.30° vertical separation (centre-to-centre), which varied because the location of all non-target stimuli (NT) was randomly jittered by ± 1.43° horizontally and vertically on each trial. The relevant distractors (relational, optimal, target-similar and salient) and the target were not jittered to retain precise eye tracking data and to ensure that these stimuli were never too close to each other. The location of all items was randomly chosen on each trial, with the restriction that the items of interest could never appear in the corner and corner-adjacent positions of the search array (as these positions were too far from fixation), and never in the central four (3.82° from fixation) positions (as these were too close to fixation).

Each coloured circle contained a letter, either 'u', 'n', 'm' or 'w' (height: 0.29°). The target stimulus always contained a 'u' or 'm'. The distractor items of interest never contained the target response characters (u/m). Colours were selected from an equiluminant (30 ± 2cd/m$^2$) RGB colour set (see Figure 2). There were two potential target colours, orange (RGB: [227, 124,52]) and teal (RGB: [56,171,146]), that alternated between blocks. The orange target could either appear among a set of redder non-targets, creating a *yellower* target condition, or among yellower non-targets, creating a *redder* target condition. The teal target could be presented either among greener or bluer non-targets, creating *bluer* or *greener* target conditions, respectively. Each trial contained the same amount of distractor and non-target items: One distractor with a relationally matching colour (R), two distractors with an optimal colour (O), two target-matching distractors (T), and a salient distractor (S) that always had a colour from the opposite side of colour space (e.g., a pink distractor for when the target was teal). The stimuli were always presented among 29 non-target items that were selected from two other colours (e.g., two blue-ish colours; see Fig. 2).

**Design.** The colours of the target, distractors and non-targets were always repeated within a mini-block of 10 trials, and mini-blocks alternated between orange and teal targets, with the direction of search (redder/yellower or greener/bluer condition) determined randomly at the start of each block. Participants completed 64 mini-blocks, for a total of 640 trials. The first four blocks were treated as practice trials, leaving 600 trials for the final analysis.

**Procedure.** Prior to the experiment, participants were instructed to locate the target-coloured stimuli containing the character *u* or *m* and respond with the corresponding keyboard key as quickly and accurately as possible. Moreover, prior to each mini-block, participants were shown the colour of the upcoming visual search target (either orange or teal). Importantly, no information about the colour of the non-targets was provided; hence participants did not know the relative colour of the target prior to the first trial (i.e., whether it was redder/yellower or greener/bluer).

To ensure stable and accurate eye tracking, participants were calibrated with a randomised 9-point calibration at the beginning of the experiment and whenever the fixation control failed. The fixation control was implemented prior to each trial, with participants maintaining fixation for 650ms on the central fixation cross. If gaze remained within 2.0° from the centre, the search array was presented until a manual keypress response (u, m) was recorded. If an incorrect response was returned, error feedback was displayed. After each trial, a blank grey screen was presented for 750ms, and the next trial commenced again with the fixation control.

## Results

Overall, accuracy on the letter identification task was high (>90%). Trials with incorrect responses were excluded from all analyses (6.6% of trials). The average response time (RT) was 2,207.7ms ($SD$ = 1,919.4ms). Trials that were more than 2.5 standard deviations above the mean RT (rounded to 7,000ms) were excluded from all analyses (1.9% of trials).

**Proportion of First Fixations on Each Item on Trials 1 - 10**

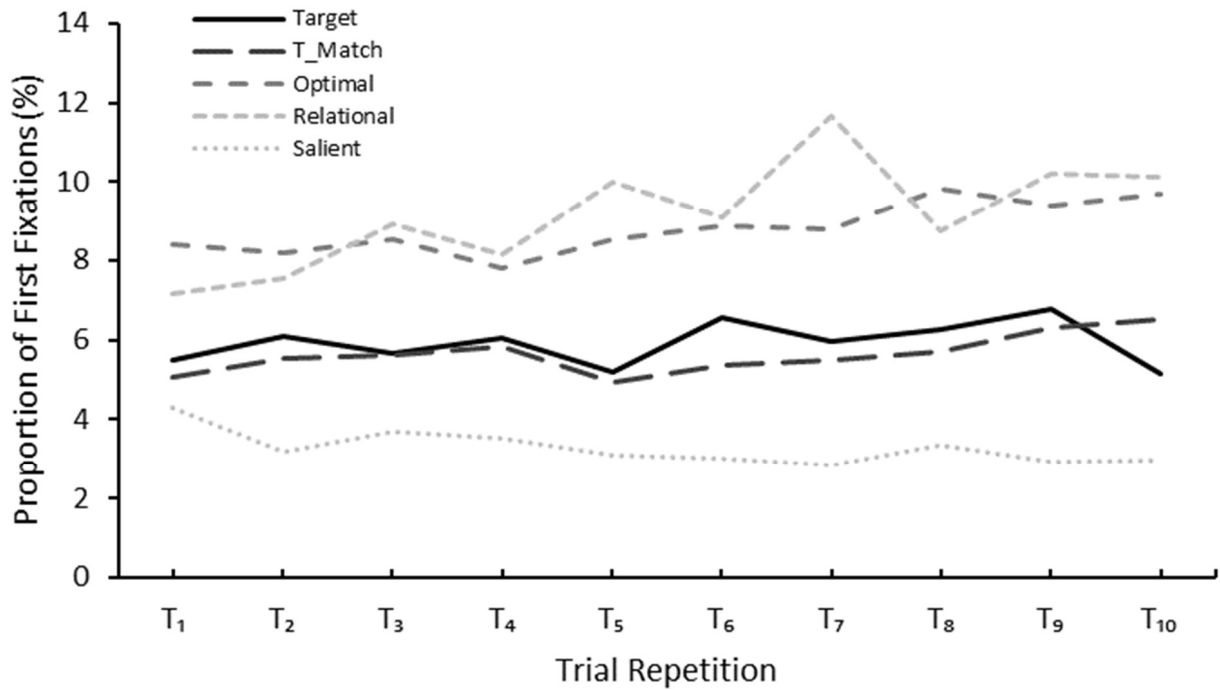

*Figure 3.* The proportion of first fixations directed towards the target, or any of the distractors of interest (target-matching, optimal, relational or salient), depicted as a function of trial repetition. The relational and optimal distractors attracted the highest proportions of first fixations and showed a linear increase over the ten trials. The target and the target-matching distractor received more first fixations than the saliently different item, and also displayed a linear increase in fixations. The salient item was selected least frequently and showed a linear decrease in first fixations.

**First Fixations: Training Effects.** To assess whether attentional priorities changed with experience, we analysed the probability of fixating on each of the differently coloured distractors as a function of trial repetition (see Fig. 3). The probability of fixating on each differently coloured distractor was computed by dividing the proportion of trials in which a given distractor was fixated (as the first item) by the number of distractors present on the trial (e.g., as there were two optimal-coloured stimuli, the proportion of first fixations on an optimal coloured stimulus was divided by two). The proportion of first fixations on the target stimulus and the target-matching distractor did not differ from each other, $F(1, 31) = 0.95$, $p = .337$, and did not interact with trial repetition, $F(9, 279) = 1.16$, $p = .321$. This shows that attention was not guided to the target item based on the contained response-related character, and thus the data were collapsed across the target and target-coloured distractor.

A 4 (Item Type: Target-Matching, Optimal, Relational, Salient) x 10 (Trial Repetition: $T_1$ to $T_{10}$) repeated measures analysis of variance (ANOVA) was computed over the probabilities of fixating on each differently coloured item. The results showed a main effect of item type, $F(3, 93) = 53.5.47, p < .001, \eta^2_p = 0.63$, trial repetition, $F(9, 279) = 4.42, p < .001, \eta^2_p = 0.13$, as well as a significant interaction, $F(27, 837) = 2.46, p < .001, \eta^2_p = 0.07$.

Pairwise $t$-tests showed that both the relational and optimal items attracted more first eye movements than the target-coloured items (relational: $t(31) = 5.33, p < .001, BF_{10} = 2500.17$, optimal: $t(31) = 6.61, p < .001, BF_{10} = 7.24 * 10^4$). In turn, the target-coloured items received more first saccades than the saliently different item, $t(31) = 3.96, p < .001, BF_{10} = 71.41$. Optimal and relational items did not differ, $t(31) = 1.07, p = .294, BF_{10} = 0.32$.

To investigate possible linear trends that may reflect learning or adaptation effects, we next examined the slopes of first fixation locations separately for each of the differently coloured items across the ten trials. The linear trends for all four item types differed significantly from zero. The relational item had the steepest positive slope, $\beta = 0.33, t(31) = 4.38, p < .001$, followed by the optimal item, $\beta = 0.18, t(31) = 2.87, p = .007$, and the target-matching item, $\beta = 0.09, t(31) = 2.36, p = .025$, reflecting that selection increased for all of these items over the course of the block. Thus, the first fixations in a trial show relational tuning, which increased slightly and remained dominant across all trials in the block. Finally, the saliently different item showed a significant reduction in visual selection across blocks, $\beta = -0.11, t(31) = 2.08, p = .046$.

**First Fixations: Initial Guidance on $T_1$.** We next analysed the eye movement behaviour on the first trial of the block, where participants were unaware of the relative colour of the target (redder/yellower or greener/bluer), and only knew the exact target colour (orange or teal; see Fig. 3, $T_1$).

Paired-samples *t*-tests revealed that first saccade proportions on $T_1$ were lower for the target colour than both the optimal, $t(31) = 4.34$, $p < .001$, $BF_{10} = 185.7$, and relational distractors, $t(31) = 2.50$, $p = .018$, $BF_{10} = 2.71$. Optimal and relational distractors did not significantly differ from each other, $t(31) = 1.59$, $p = .123$, $BF_{10} = 0.58$. The salient item did not differ significantly from the target-coloured item, $t(31) = 0.88$, $p = .388$, $BF_{10} = 0.27$. The finding that relational search was conducted on trial 1 shows that experience or prior knowledge is not necessary for relational search, and supports the hypothesis that the visual system can quickly assess the dominant colour in the visual field and tune attention to the relative colour of the target.

**Fixation Progression.**

The results above suggest that search was initially (on the first trial) relational and that this relational guidance was maintained and increased across repeated trials. To assess whether the same trends may be found *within a single trial*, we analysed the first five fixations in each trial to reveal how attentional guidance developed within an individual trial (see Figure 4). On average participants made 4.58 fixations per trial. We included all trials in this data set, including those where the task was completed within the first five fixations (which resulted in 1.7% missing data for $F_2$, $F_3$: 7.4%, $F_4$: 17.5%, and $F_5$: 30.6%).

The first fixation ($F_1$), now collapsed over trial repetition, followed the same results pattern as previously described, with the relational and optimal stimuli attracting a higher proportion of first fixations than the target and the target-matching distractor (all $p$s < .001). For $F_2$ results began to deviate from relational guidance. The target stimulus was now more likely to be fixated than any other item ($p$s < .010). The target-matching distractors were now equally likely to be fixated as the optimal distractors, $t(31) = 0.14$, $p = .888$, $BF_{10} = 0.19$, and the optimal distractors were now more likely to be fixated than the relational item, $t(31) =$ 4.32, $p < .001$, $BF_{10} = 178.09$. From $F_3$ onwards fixation patterns remained consistent: Now the target-matching distractors were more likely to be fixated than the optimal distractor (all $p$s $< .001$), and the optimal distractor was more likely to be fixated than the relational distractor (all $p$s $< .001$). Finally, fixations on the salient item were highest on $F_1$ than at any other point in the trial (all $p$s $< .001$) and were significantly lower than all other stimuli of interest across all fixations ($p$s $< .004$).

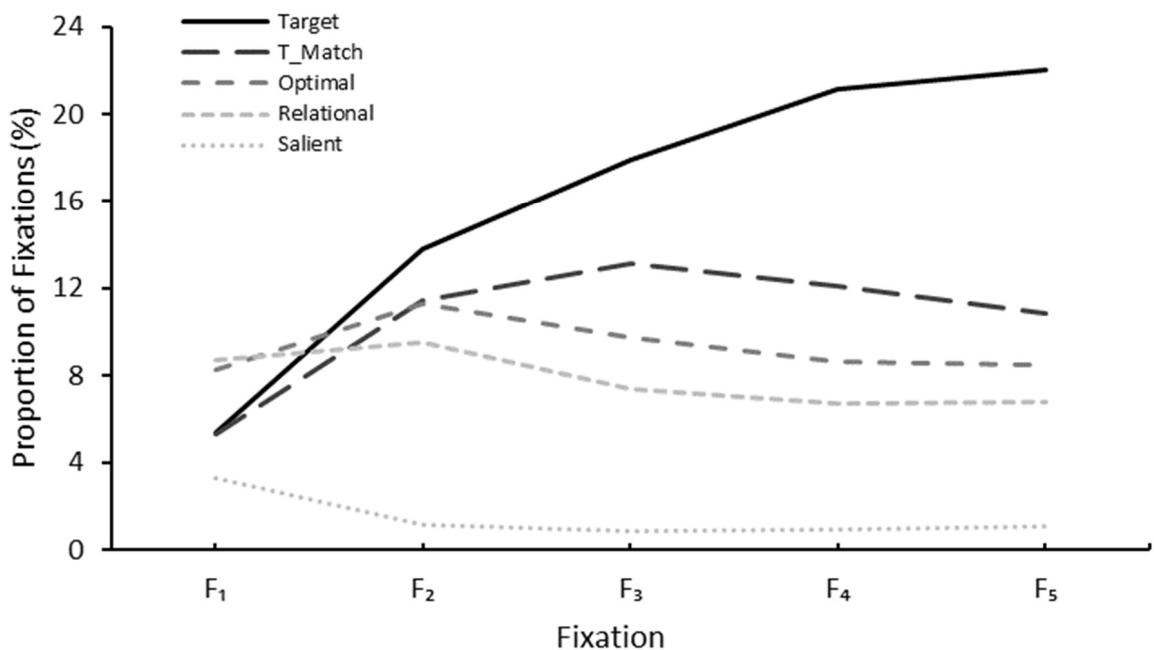

*Figure 4.* The first five fixations ($F_1$ to $F_5$) in a trial show how search progressed during a trial. The first fixation ($F_1$) displayed relational search, with the relational and optimal distractors being selected more frequently than the target and target-matching distractor. The second fixation ($F_2$) showed a shift to more feature-specific guidance: The target was now the most likely to be fixated, followed by the target-matching and optimal distractors, with fewer fixations on the relational distractor. With the third fixation ($F_3$), the target-matching distractor was now more likely to be fixated than optimal or relational distractors. From this point onwards, fixation proportions seem to be dictated by target-feature similarity. The salient item was most frequently fixated on $F_1$ then dropped off in subsequent fixations.

In sum, the results are in line with the hypothesis that (in relational search) participants will first select the relationally maximal item (e.g., reddest), followed by the optimal (e.g., next-reddest) item, and the target-coloured items. Another way of describing these results is that attention was always initially guided to the relatively matching items, with feature-specific tuning developing only after the first (few) fixation(s), perhaps by inhibiting the more extreme colours that were selected first.

**First Saccade Latencies.** We next analysed the saccade latencies, that is, the time from the onset of the search display to the onset of the first saccade in a trial, to assess the time-course of attentional deployment using the first saccade in a trial. This analysis was collapsed over trials, with the first trial being excluded (due to longer saccadic latencies).[1] Figure 5 shows the distributions of saccadic latencies separately for each item type (i.e., the proportion of trials where saccadic latencies ranged from 125ms – 150ms, 150 – 175ms, etc.).
To analyse the data statistically, we fed the average saccade latencies of the first saccades in a trial into a one-way (Item Type: Target, Optimal, Relational, Salient) repeated-measures ANOVA. The results showed a significant effect of item type $F(3,93) = 53.01$, $p < .001$, $\eta^2_p = 0.63$. Two-tailed comparisons revealed that latencies were shorter for the salient item ($M = 217.6$ms) than both the relational ($M = 237.5$ms) and optimal item ($M = 240.9$ms), all $t$s > 8.18, $p$s < .001, $BF_{10} > 3.98 \times 10^6$. Saccades to target-matching items ($M = 252.0$ms) were slower than to the relational and optimal items, $t$s(31) = 4.35, $p$s < .001, $BF_{10} > 191.72$, while relational and optimal items did not differ, $t(31) = 1.26$, $p = .218$, $BF_{10} = 0.39$. Thus, the more dissimilar the stimuli were from the target item, the shorter their saccadic latencies were.

---

[1] Saccade latencies on $T_1$ were longer than in all other trial repetitions (all $p$s < .001), but there was no repetition by item type interaction ($p = .650$) or any linear trends for the individual items after excluding the first trial ($p$s > .235).

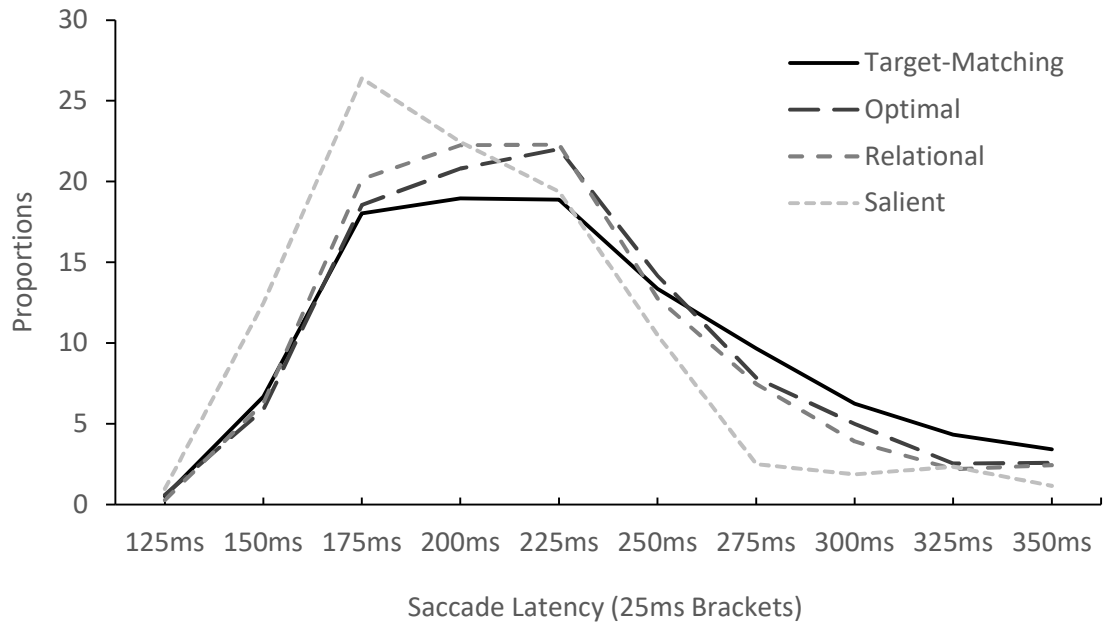

*Figure 5.* The distribution of saccade latencies, which was derived by sorting saccades to each stimulus type into 25ms latency bins and depicting the proportions of trials within each bin separately for each item type. The results showed a higher proportion of short-latency saccades to the saliently different distractor than to the other distractors. Moreover, there was a higher proportion of saccades with long latencies to target-matching items (pooled over target and target-coloured distractor), producing differences between the target-coloured items and the other distractors in the tail end of the distribution.

Figure 5 shows that the distribution of saccadic latencies and reveals that differences in the average saccade latencies were due to saccades to salient distractor being initiated significantly earlier than to all other items. By contrast, the longer saccade latencies for the target and target-matching distractors were due to a higher proportion of saccades with longer latencies.

**Distractor Dwell Times.** As in previous studies, we also examined the mean dwell times on each of the distractors, measured as the time spent fixating on each of the different item types. The target-matching stimuli were examined independently from the actual target ($M =$ 162.9ms), and results were collapsed over trial repetition.

The one-way (Distractor Type: Target-Matching, Optimal, Relational, Salient) repeated-measures ANOVA revealed significant differences between the item types, $F(3,93) = 40.50$, $p < .001$, $\eta^2_p = 0.57$. Planned two-tailed comparisons showed that dwell times for the target-matching ($M = 173.0$ms) and optimal distractors ($M = 171.4$ms) were longest and did not differ from each other, $t(31) = 0.92$, $p = .363$, $BF_{10} = 0.28$. Dwell times were significantly shorter for the relational distractor ($M = 166.4$ms) than the optimal distractor, $t(31) = 5.74$, $p < .001$, $BF_{10} = 7312.78$, and target matching distractors, $t(31) = 3.21$, $p = .003$, $BF_{10} = 11.99$. The relational distractor had longer dwell times than the salient item ($M = 152.5$ms), $t(31) = 6.26$, $p < .001$, $BF_{10} = 2.90 * 10^4$. In sum, dwell times increased with similarity to the target colour, indicating that longer dwell times were needed to identify the items as distractors as they became more similar to the target.

## General Discussion

The current study revealed that relational effects may be more pervasive than previously assumed. This was the first study to examine attentional tuning functions in the absence of target contextual knowledge and in displays containing numerous items. The relatively matching distractors attracted the first saccade, including on the *first* trial in a mini-block, when participants only knew the colour of the target but not its relative contextual colour. These results show that relational/contextual information can be quickly extracted from a visual scene, in line with the tenets of the relational account. Previous studies have already shown that information about the statistical properties of a visual scene can be rapidly extracted (e.g., feature averaging, Chong & Treisman, 2005a, b; Choo & Franconeri, 2010; Joo et al., 2009; gist processing, Oliva & Torralba, 2006; see also Thorpe, Fize & Marlot, 1996). The present study extends on this research by showing that this information can be used to bias attention to the relative features of the target, prior to executing the first eye movement. In other words, relational guidance does not require learning of the context or knowledge of the relative target feature, but can be executed with the first glance at a novel scene.

The current results also revealed that relational guidance is more ubiquitous than previously thought. In particular, we found relational guidance even when the target never had a relationally maximal value – that is, when it was never the reddest, greenest, bluest or yellowest item in the display. In every trial there were relationally better-matching distractors (one relatively matching distractor and two optimal distractors), which frustrated selecting the target with the first eye movement (in relational search). Yet, we reliably found that participants searched relationally. This reveals that the visual system is indeed sensitive to the dominant feature in the display, and initiates relational search as soon as the target differs from *most* of the display items in a linear fashion. Furthermore, participants continued to search relationally over as many as ten successive trials, even though relational guidance reliably resulted in selecting one of the more extreme distractors first. This means that relational guidance is more readily applied and more persistent than previously thought (e.g., Becker et al., 2014).

The results of the present study also provided evidence for selection proceeding in the way as laid out in the relational account, with the most extreme, relatively matching item being selected first (e.g., reddest), followed by the next most extreme, relatively matching item (i.e., next reddest item), and so forth, until the target is selected. As shown in Figure 3, the first saccade in a trial was most likely to select the most extreme, relatively matching item, whereas later saccades were more likely to select target-matching items and the probability for selecting extreme items declined after the second fixation.

The mechanism responsible for achieving this order of selection has never been specified in the relational account. Other theories have proposed a memory-based, 'inhibitory tagging' mechanism to explain how search proceeds to the next item (e.g., Klein, 2000).

Correspondingly, it is possible that selection of target-matching items emerged during the trial due to selection and subsequent inhibition of relatively matching, more extreme distractors. Inhibitory tagging could be either location-based (i.e., inhibiting selected nontarget locations during the trial), or feature-based (i.e., inhibiting a non-target or distractor colour after selection; e.g., Bichot & Schall, 2002). If inhibitory tagging is responsible for the selection sequence in the present study, it is perhaps more likely to be feature-based inhibition rather than location-based inhibition: Selection of the target and target-matching distractors commenced earlier than would be expected on the basis of location-based inhibition. Specifically, as there were three relatively matching or optimal distractors in the display, selection of these distractors should have declined after the $3^{rd}$ fixation, but we observed the decline after the $2^{nd}$ fixation, which corresponds to the number of different colours in these distractors. Also, selection of the target and target-matching items showed a steep increase after only one fixation, which seems too early for location-based inhibition, and indicates that the colour(s) rather than the locations of the more extreme distractors may have been inhibited.

Alternatively, it is possible that, after the first fixation, attention was tuned to the exact feature value of the target, swiftly switching from originally relational search to a more feature-specific search (or target template; e.g., Duncan & Humphreys, 1989). We currently consider this unlikely, as an attentional bias usually automatically carries over to the next trial to influence selection (i.e., intertrial priming; Maljkovic & Nakayama, 1994), and the first eye movements never showed evidence of a feature-specific bias.

Inhibition of a distractor feature can also carry over to the next trial to influence selection, but the effects are weaker than effects of target guidance (e.g., Chang & Egeth, 2019; Hamblin-Frohman et al., 2022). This could explain that relational guidance was somewhat attenuated, as the relatively matching distractor mostly did not differ significantly from the optimal distractor, but was still clearly visible (as both the relatively matching and optimal distractors were selected more frequently than the target-matching distractors; see Fig. 3). However, the somewhat attenuated effect of the relatively matching distractor could also be due to difficulties distinguishing the optimal and relatively matching distractor in the far periphery (e.g., Noorlander et al., 1983), or to the fact that there was only one relatively matching, extreme distractor and two optimal distractors. Further studies are necessary to estimate the magnitude of the attenuation and clarify the mechanism that allows narrowing search to target-matching items after the first fixation(s).

A fourth interesting finding relates to the salient item, which had a very dissimilar colour compared to the other items and hence, had the largest bottom-up feature contrast. Wang and Theeuwes (2020) proposed that experiments with sparse displays may underestimate bottom-up saliency effects (see also Rangelov et al., 2017). The displays of the present study contained the largest number of items tested for fine-grained attentional tuning (36 items), yet we found only very weak effects for the salient item, which further declined over trials 1 – 10 (see Fig. 3). Given the low proportion of first fixations on the salient item, we cannot claim that the salient item significantly attracted attention.[2] Even in the densely populated displays, top-down tuning to the (relative) target feature led to significantly higher selection rates of the corresponding distractors than bottom-up saliency, indicating that top-down processes dominate attentional guidance over bottom-up saliency, contrary to bottom-up selection views (e.g., Theeuwes, 2004; Wang & Theeuwes, 2020).

These results also rule out an alternative explanation of the results. Proponents of feature-based theories (e.g., Guided Search 2.0; Wolfe, 1994) may argue that the results could be due to a combination of top-down, feature-specific tuning to the target and bottom-up processes, as the more extreme distractors (optimal and relatively matching distractors) may have been similar enough to the target to be subject to (broad) top-down tuning and were simultaneously more salient than the target, leading to higher selection rates. Contrary to this contention, we found a very different results pattern for the salient distractor compared to the other distractors, both in the trial analysis (which showed an increase in capture by extreme and target-similar distractors over trials and a decrease for the salient distractor; see Fig. 3), and in the time-course analysis (which showed shorter latencies of saccades to the salient distractor than for all other distractors; see Fig. 5). These findings suggest that the more extreme (and more salient) relatively matching and optimal distractors attracted attention and the gaze in the same manner as the target-similar (less salient) distractor, viz., due to top-down tuning and not bottom-up saliency, whereas the effects of the salient distractor were due to bottom-up, stimulus-driven processes.

---

[2] Comparing the proportion of first fixations on the salient item to the proportion of fixations on the non-salient non-target items (omitting the corner positions and central positions that never contained a distractor) revealed that the salient item was actually selected *less* frequently than the non-salient non-targets (*p*=.03).

Admittedly, it is notoriously difficult to distinguish between combined top-down/bottom-up theories, optimal tuning and the relational account, especially in standard visual search experiments with a distractor, as the theories make quite similar predictions. However, the relational account has been extensively tested against these other theories in previous experiments using the spatial cueing paradigm (e.g., Becker et al., 2013) and visual search with relatively matching distractors that were very dissimilar from the target (York & Becker, 2020) or whose similarity / saliency varied systematically (Hamblin-Frohman & Becker, 2021; Yu et al., 2022). The results unequivocally supported the relational account and showed that early selection cannot be explained by optimal tuning or the combined saliency and similarity to the target.

Previous studies testing the relational account against the standard, feature-based theories moreover revealed that tuning to relations was a *default* search mode applied across a wide range of different tasks that would have allowed locating the target by tuning to its specific feature value (e.g., when the target was always repeated). The present study extends on these findings, by showing that attention is tuned to relative features even when the target is never the relatively reddest, bluest, greenest or yellowest item in the display. In conditions such as the present experiment, a feature-specific guidance mechanism may be the most efficient search method for the current display, as it would have reduced competition for attention to only target-similar distractors (i.e., two other items) by excluding relatively matching and optimal distractors. Participants chose relational tuning over feature-specific tuning even though, on the first trial of each block, they were only given information about the exact feature-value of the target stimulus. The instructional prompts mentioning only the target colour should have primed participants to initiate a guided search for the exact target colour. Yet, participants consistently engaged in relational search and also did not learn to use feature-specific tuning over the course of ten (identical) trials. As shown in Figure 3, linear trends for first eye movements were stronger for the relational and the optimal distractors compared to the target-matching items, suggesting that the relational effect became *stronger* over the course of the block rather than weaker.

Collectively, these results show that participants can be quite reluctant to tune attention to the specific feature value of the target. We can only speculate why this may be the case. Previous studies have shown that feature-specific tuning results in delays in selecting the target, as well as longer dwell times on target-similar distractors, compared to tuning to relations (in identical displays; Becker et al., 2014; Martin & Becker, 2018). This suggests that tuning attention to exact feature values may cause delays in selection as well as decision-making (about whether the selected item is the target), rendering search less efficient.

While this question warrants further research, the present study clearly showed that tuning to relative features occurs in more densely populated, 36-item displays; in the absence of knowledge of the relative feature of the target, and in the absence of any training. This demonstrates the existence of a mechanism capable of extracting the dominant feature in a visual scene and using that information to tune attention to the relative feature of a target. This renders it likely that relational search is the default search mode in everyday situations.

**Author note.**

This research was supported by Australian Research Council (ARC) grant DP210103430 to SIB.
27

**References**

Becker, S. I. (2010). The role of target-distractor relationships in guiding attention and the eyes in visual search. Journal of Experimental Psychology. General, 139(2), 247–265. https://doi.org/10.1037/a0018808

Becker, S. I. (2013). Simply shapely: Relative, not absolute shapes are primed in pop-out search. Attention, Perception, and Psychophysics, 75(5), 845–861. https://doi.org/10.3758/S13414-013-0433-1/FIGURES/4

Becker, S. I., Folk, C. L., & Remington, R. W. (2013). Attentional Capture Does Not Depend on Feature Similarity, but on Target-Nontarget Relations. Psychological Science, 24(5), 634–647. https://doi.org/10.1177/0956797612458528

Becker, S. I., Harris, A. M., Venini, D., & Retell, J. D. (2014). Visual search for color and shape: When is the gaze guided by feature relationships, when by feature values? Journal of Experimental Psychology: Human Perception and Performance, 40(1), 264–291. https://doi.org/10.1037/a0033489

Becker, S. I., Harris, A. M., York, A., & Choi, J. (2017). Conjunction search is relational: Behavioral and electrophysiological evidence. Journal of Experimental Psychology: Human Perception and Performance, 43(10), 1828. https://doi.org/10.1037/XHP0000371

Becker, S. I., Valuch, C., & Ansorge, U. (2014). Color priming in pop-out search depends on the relative color of the target. Frontiers in Psychology, 5. https://doi.org/10.3389/fpsyg.2014.00289

Bichot, N. P., & Schall, J. D. (2002). Priming in Macaque Frontal Cortex during Popout Visual Search: Feature-Based Facilitation and Location-Based Inhibition of Return. Journal of Neuroscience, 22(11), 4675–4685. https://doi.org/10.1523/JNEUROSCI.22-11-04675.2002

Carrasco, M. (2011). Visual attention: The past 25 years. Vision Research, 51(13), 1484–1525. https://doi.org/10.1016/J.VISRES.2011.04.012

Chang, S., & Egeth, H. E. (2019). Enhancement and Suppression Flexibly Guide Attention. Psychological Science, 30(12), 1724–1732. https://doi.org/10.1177/0956797619878813

Chong, S. C., & Treisman, A. (2005a). Attentional spread in the statistical processing of visual displays. Perception & Psychophysics, 67(1), 1–13. https://doi.org/10.3758/BF03195009

Chong, S. C., & Treisman, A. (2005b). Statistical processing: computing the average size in perceptual groups. Vision Research, 45(7), 891–900. https://doi.org/10.1016/j.visres.2004.10.004

Choo, H., & Franconeri, S. L. (2010). Objects with reduced visibility still contribute to size averaging. Attention, Perception, & Psychophysics, 72(1), 86–99. https://doi.org/10.3758/APP.72.1.86

Desimone, R., & Duncan, J. (1995). Neural Mechanisms of Selective Visual Attention. Annual Review of Neuroscience, 18(1), 193–222. https://doi.org/10.1146/annurev.ne.18.030195.001205

Deubel, H., & Schneider, W. X. (1996). Saccade target selection and object recognition: Evidence for a common attentional mechanism. Vision Research, 36(12), 1827–1837. https://doi.org/10.1016/0042-6989(95)00294-4

Duncan, J., & Humphreys, G. W. (1989). Visual search and stimulus similarity. Psychological Review, 96(3), 433–458. https://doi.org/10.1037/0033-295X.96.3.433

Gaspelin, N., Gaspar, J. M., & Luck, S. J. (2019). Oculomotor inhibition of salient distractors: Voluntary inhibition cannot override selection history. Visual Cognition, 27(3–4), 227–246. https://doi.org/10.1080/13506285.2019.1600090

Gaspelin, N., Leonard, C. J., & Luck, S. J. (2015). Direct Evidence for Active Suppression of Salient-but-Irrelevant Sensory Inputs. Psychological Science, 26(11), 1740–1750. https://doi.org/10.1177/0956797615597913

Gaspelin, N., & Luck, S. J. (2018). The Role of Inhibition in Avoiding Distraction by Salient Stimuli. Trends in Cognitive Sciences, 22(1), 79–92. https://doi.org/10.1016/J.TICS.2017.11.001

Gaspelin, N., Margett-Jordan, T., Ruthruff, E., Gaspelin, N., Margett-Jordan, T., & Ruthruff, E. (2015). Susceptible to distraction: Children lack top-down control over spatial attention capture. Psychon Bull Rev, 22, 461–468. https://doi.org/10.3758/s13423-014-0708-0

Geng, J. J., di Quattro, N. E., & Helm, J. (2017). Distractor probability changes the shape of the attentional template. Journal of Experimental Psychology: Human Perception and Performance, 43(12), 1993. https://doi.org/10.1037/XHP0000430

Hamblin-Frohman, Z., & Becker, S. I. (2021). The attentional template in high and low similarity search: Optimal tuning or tuning to relations? Cognition, 212, 104732. https://doi.org/10.1016/j.cognition.2021.104732

Hamblin-Frohman, Z., Chang, S., Egeth, H., & Becker, S. I. (2022). Eye movements reveal the contributions of early and late processes of enhancement and suppression to the guidance of visual search. Attention, Perception, and Psychophysics, 84(6), 1913–1924. https://doi.org/10.3758/S13414-022-02536-W/FIGURES/4

Joo, S. J., Shin, K., Chong, S. C., & Blake, R. (2009). On the nature of the stimulus information necessary for estimating mean size of visual arrays. Journal of Vision, 9(9), 7–7. https://doi.org/10.1167/9.9.7

Klein, R. M. (2000). Inhibition of return. Trends in Cognitive Sciences, 4(4), 138–147. https://doi.org/10.1016/S1364-6613(00)01452-2

Maljkovic, V., & Nakayama, K. (1994). Priming of pop-out: I. Role of features. Memory & Cognition, 22(6), 657–672. https://doi.org/10.3758/BF03209251/METRICS

Martin, A., & Becker, S. I. (2018). How feature relationships influence attention and awareness: Evidence from eye movements and EEG. Journal of Experimental Psychology: Human Perception and Performance, 44(12), 1865. https://doi.org/10.1037/XHP0000574

Navalpakkam, V., & Itti, L. (2007). Search Goal Tunes Visual Features Optimally. Neuron, 53(4), 605–617. https://doi.org/10.1016/J.NEURON.2007.01.018

Noorlander, C., Koenderink, J. J., den Olden, R. J., & Edens, B. W. (1983). Sensitivity to spatiotemporal colour contrast in the peripheral visual field. Vision Research, 23(1), 1–11. https://doi.org/10.1016/0042-6989(83)90035-4

Oliva, A., & Torralba, A. (2006). Chapter 2 Building the gist of a scene: the role of global image features in recognition (pp. 23–36). https://doi.org/10.1016/S0079-6123(06)55002-2

Peirce, J. W. (2007). PsychoPy—Psychophysics software in Python. Journal of Neuroscience Methods, 162(1–2), 8–13. https://doi.org/10.1016/J.JNEUMETH.2006.11.017

Schönhammer, J. G., Grubert, A., Kerzel, D., & Becker, S. I. (2016). Attentional guidance by relative features: Behavioral and electrophysiological evidence. Psychophysiology, 53(7), 1074–1083. https://doi.org/10.1111/psyp.12645

Scolari, M., Byers, A., & Serences, J. T. (2012). Optimal Deployment of Attentional Gain during Fine Discriminations. Journal of Neuroscience, 32(22), 7723–7733. https://doi.org/10.1523/JNEUROSCI.5558-11.2012

Theeuwes, J. (2004). Top-down search strategies cannot override attentional capture. Psychonomic Bulletin & Review, 11(1), 65–70. https://doi.org/10.3758/BF03206462

Theeuwes, J. (2013). Feature-based attention: it is all bottom-up priming. Philosophical Transactions of the Royal Society B: Biological Sciences, 368(1628). https://doi.org/10.1098/RSTB.2013.0055

Thorpe, S., Fize, D., & Marlot, C. (1996). Speed of processing in the human visual system. Nature, 381(6582), 520–522. https://doi.org/10.1038/381520a0

Wang, B., & Theeuwes, J. (2020). Salience determines attentional orienting in visual selection. Journal of Experimental Psychology: Human Perception and Performance, 46(10), 1051–1057. https://doi.org/10.1037/xhp0000796

Wolfe, J. M. (1994). Guided Search 2.0 A revised model of visual search. Psychonomic Bulletin & Review, 1(2), 202–238. https://doi.org/10.3758/BF03200774/METRICS

Wolfe, J. M. (2020). Visual Search: How Do We Find What We Are Looking For? Annual Review of Vision Science, 6(1), 539–562. https://doi.org/10.1146/annurev-vision-091718-015048

Wolfe, J. M. (2021). Guided Search 6.0: An updated model of visual search. Psychonomic Bulletin & Review 2021 28:4, 28(4), 1060–1092. https://doi.org/10.3758/S13423-020-01859-9

York, A., & Becker, S. I. (2020). Top-down modulation of gaze capture: Feature similarity, optimal tuning, or tuning to relative features? Journal of Vision, 20(4), 6–6. https://doi.org/10.1167/JOV.20.4.6

Yu, X., Hanks, T. D., & Geng, J. J. (2022). Attentional Guidance and Match Decisions Rely on Different Template Information During Visual Search. Psychological Science, 33(1), 105–120. https://doi.org/10.1177/09567976211032225/ASSET/IMAGES/LARGE/10.1177_09567976211032225-FIG2.JPEG